\documentclass[preprint,aps,superscriptaddress,nofootinbib,showpacs]{revtex4}
\usepackage{amsmath,amssymb}
\usepackage{graphicx,color,subfig}
\usepackage{slashed}

\begin{document}

\title{A Holographic Model of Two-Band Superconductor}
\author{Wen-Yu Wen}
\email{steve.wen@gmail.com}
\affiliation{Department of Physics and Chung Yuan Center for High Energy Physics, Chung Yuan Christian University, Chung Li City, Taiwan}
\affiliation{Leung Center for Cosmology and Particle Astrophysics\\
National Taiwan University, Taipei 106, Taiwan}
\author{Mu-Sheng Wu}
\email{msgn123@gmail.com}
\affiliation{Physics Division, National Center for Theoretical Sciences, Hsinchu 300, Taiwan}
\affiliation{Department of Physics, National Tsing Hua University, Hsinchu 300, Taiwan}
\author{Shang-Yu Wu}
\email{loganwu@gmail.com}
\affiliation{Institute of physics, and Shing-Tung Yau Center, National Chiao Tung University, Hsinchu 300, Taiwan}
\affiliation{National Center for Theoretical Science, Hsinchu 300, Taiwan}
\begin{abstract}
We construct a holographic two-band superconductor model with interband Josephson coupling. We investigate the effects the Josephson coupling has on the superconducting condensates and the critical temperature for their formation numerically, as well as analytically where possible. We calculate the AC conductivity and find it qualitatively similar to the single band superconductor. We investigate the nodal structure of our holographic two-band superconductor from the low temperature behavior of the thermal conductivity and find it nodeless.

\end{abstract}

\pacs{11.25.Tq, 74.20.-z}
\maketitle


\section{Introduction}

The AdS/CFT correspondence~\cite{Maldacena, Polyakov, Witten} has proved very useful in providing novel tools to study strongly-coupled/correlated systems. It has been applied to, e.g. RHIC physics~\cite{Chernicoff:2012bu,Yee:2013qma,Wu:2013qja,v_2,DeWolfe:2013cua}, and recently to condensed matter phenomena~\cite{Herzog:2007ij,Hartnoll:2007ih,Hartnoll:2007ip,Hartnoll:2008hs,Minic:2008an,Sun:2013wpa,Sun:2013zga} (for a review, see e.g. Ref.~\cite{Hartnoll:2009sz}). A gravity model was proposed in Refs.~\cite{Gubser:2005ih,Gubser:2008px} in which a $U(1)$ symmetry is spontaneously broken by the existence of a black hole. This mechanism was recently incorporated in the model of superconductivity: critical temperature and magnetic field were observed~\cite{Hartnoll:2008vx,Nakano:2008xc,Albash:2008eh}\footnote{The issue of emergent dynamical gauge field in holographic superconductor is discussed in \cite{Domenech:2010nf}.}, and later non-Abelian gauge condensate~\cite{Gubser:2008zu} and condensate of higher spin~\cite{Gubser:2008wv,
Roberts:2008ns,Chen:2010mk}. Some interesting phenomena observed in the laboratory also appeared in the study of fermion
spectral functions~\cite{Chen:2009pt,Benini:2010qc,Chen:2011ny}.

Historically, Ginzburg-Landau theory has proved to be an extraordinarily valuable phenomenological tool in understanding single-component superconductors. Its generalization to the two-component Ginzburg-Landau model (TCGL) was constructed, and its applicability to the two-band systems studied in Refs.~\cite{Silaev:2012,Shanenko:2011,Vagov:2012}. Upon switching on the interband coupling between the two components, this model can describe the phenomenon of the two gaps in materials such as MgB$_{2}$ ($s_{++}$)\cite{Carlstrom:2010,Buzea:2001} and iron pnictides ($s_{+-}$)\footnote{Another mechanism due to the shape resonance is also proposed to explain some cases of the iron pnictides \cite{Bianconi:2013}. We thank to Prof. Antonio Bianconi for bringing this interesting reference.}\cite{iron-based1,iron-based2,iron-based3}. A holographic model with two order parameters was first studied in the probe limit~\cite{Basu:2010fa}, and recently with back-reaction~\cite{Cai:2013wma}, where phases with two condensates
coexisting and competing were observed.  However, the absence of an interband (Josephson) coupling in those models makes it difficult to justify them as models of two-band superconductivity where the interaction between the two bands is crucial. A multi-band holographic model
for three coherent orders was discussed in Ref.~\cite{Huang:2011ac}. However, in their model the form and strength of the interband interaction is completely fixed by the built-in $SO(3)$ gauge symmetry in the bulk, and is not a parameter that can be tuned. A similar holographic model for the two-band case based on an $U(2)$ symmetry was also constructed~\cite{Krikun:2012yj} in which the two condensates can be of the same or opposite sign, i.e. zero or $\pi$ relative phase difference.

In this paper, we study the effects the interband coupling has on the superconducting condensates and the critical temperature of their formation in a holographic model adapted from that proposed in Ref.~\cite{Aprile:2009ai}, which has a tunable interband Josephson coupling. In the language of the TCGL model, for positive Josephson coupling the two-band superconductor is in the same sign, $s_{++}$, state, while for negative Josephson coupling, it is in the opposite sign, $s_{+-}$, state. A defining characteristic of the two-band superconductor is the existence of coherent orders in which the two orders have the same critical temperature. Here we look for this characteristic feature in our holographic two-band superconductor, and we study it electrical and thermal transport properties. The thermal conductivity, $\kappa$, is of particular interest as its low temperature behavior provides a good probe of the superconducting gap structure experimentally. The contribution to the thermal conductivity due to
conduction electrons is expected to behaves as $\sim T$ at low temperatures, while that due to phonons $\sim T^3$. Thus a linear temperature dependence in $\kappa$ at low temperatures may be attributed to electron excitations. Now $\kappa \rightarrow 0$ as
$T \rightarrow 0$ would point to a fully gapped superconductor, but a finite value can indicate either a nodal structure due to pairing symmetry, or strong electron-electron interactions, or gapless behavior due to scattering.

The paper is organized as follows. We describe our holographic two-band model in Section~II. Results from our numerical study of the condensates and the electric and thermal conductivities are reported in the Section~III. We derive analytical results in regimes where it is possible in Section~IV. We end with a summary and directions for the future in Section~V.

\section{The Model}

We start by putting a generalized two-component Ginzburg-Landau theory into the (3+1)-dimensional Einstein-Maxwell-Dilaton gravity:
\begin{equation}
2\kappa_G^2 (-g)^{-1/2}{\cal L} = R + \frac{6}{L^2} - \frac{G(\varphi_{1},\varphi_{2})}{4} F^{\mu\nu}F_{\mu\nu} -\frac{1}{2}|D_{\mu}\varphi_{1}|^2 - \frac{1}{2}|D_{\mu}\varphi_{2}|^2 - V(\varphi_{1},\varphi_{2}),
\end{equation}
where $G(\varphi_{1},\varphi_{2})=1+\kappa_{1}\varphi^{\ast}_{1}\varphi_{1}+\kappa_{2}\varphi^{\ast}_{2}\varphi_{2}$ is the non-minimal coupling between the charged scalars and gauge field, and $\varphi_{1}$ and $\varphi_{2}$ are charged scalars. Except for the mass terms for the two charged scalars, we also introduce the interactions between the two charged scalars in the potential term
\begin{equation}
V(\varphi_{1},\varphi_{2}) = m^{2}_{1}\varphi^{\ast}_{1}\varphi_{1} + m^{2}_{2}\varphi^{\ast}_{2}\varphi_{2} + \epsilon (\varphi_{1}^{\ast}\varphi_{2} + \varphi_{1}\varphi_{2}^{\ast}) + \eta |\varphi_{1}|^2|\varphi_{2}|^2,
\end{equation}
where the $\epsilon$ term is the Josephson coupling introduced in the field theory literature, and the last term is the direct coupling~\cite{Basu:2010fa}. Since $\varphi_{1}$, $\varphi_{2}$ are complex scalars, we may parameterize them as $\varphi_{1}=\psi_{1}e^{i\theta_{1}}$, $\varphi_{2}=\psi_{2}e^{i\theta_{2}}$.  Then the bulk action can be rewritten as
\begin{eqnarray}
S&=&\frac{1}{2\kappa^{2}_{G}}\int{d^{4}x}\sqrt{-g}[R+\frac{6}{L^{2}}-\frac{1}{4}G(\psi_{1},\psi_{2})F_{\mu\nu}F^{\mu\nu}-\frac{1}{2}(\partial\psi_{1})^{2}-\frac{1}{2}\psi^{2}_{1}(\partial_{\mu}\theta_{1}-A_{\mu})^{2}\nonumber\\
&&-\frac{1}{2}(\partial\psi_{2})^{2}-\frac{1}{2}\psi^{2}_{2}(\partial_{\mu}\theta_{2}-A_{\mu})^{2}-V(\psi_{1},\psi_{2})], \nonumber
\end{eqnarray}
which is invariant under the gauge transformation
\begin{equation*}
A_{\mu}\rightarrow A_{\mu}+\partial_{\mu}\alpha, \quad \theta_{1}\rightarrow\theta_{1}+\alpha, \quad \theta_{2}\rightarrow\theta_{2}+\alpha. \nonumber
\end{equation*}
To preserve the gauge transformation, we can generalize the action as \cite{Aprile:2009ai}
\begin{eqnarray}
S&=&\frac{1}{2\kappa^{2}_{G}}\int{d^{4}x}\sqrt{-g}[R+\frac{6}{L^{2}}-\frac{1}{4}G(\psi_{1},\psi_{2})F_{\mu\nu}F^{\mu\nu}-\frac{1}{2}(\partial\psi_{1})^{2}-\frac{1}{2}J_{1}(\psi_{1})(\partial_{\mu}\theta_{1}-A_{\mu})^{2}\nonumber\\
&&-\frac{1}{2}(\partial\psi_{2})^{2}-\frac{1}{2}J_{2}(\psi_{2})(\partial_{\mu}\theta_{2}-A_{\mu})^{2}-V(\psi_{1},\psi_{2})],
\end{eqnarray}
where $J_{1}(\psi_{1})$, $J_{2}(\phi_{2})$ are arbitrary functions of $\psi_{1}$, $\psi_{2}$, and
\begin{eqnarray}
&&G(\psi_{1},\psi_{2})=1+\kappa_{1}\psi_{1}^{2}+\kappa_{2}\psi_{2}^{2}, \nonumber\\
&&V(\psi_{1},\psi_{2})=m_{1}^{2}\psi_{1}^{2}+m_{2}^{2}\psi_{2}^{2}+2\epsilon\psi_{1}\psi_{2}+\eta\psi_{1}^{2}\psi_{2}^{2}.
\end{eqnarray}
In the following we only consider the minimal model which gives the phase-locking condition, saying $\theta_{1}=\theta_{2}\equiv\theta$~\cite{phase-locking}. Since we do not consider the vortex solution, we can consistently set $\theta$ to be any constant, say $\theta=0$ for simplicity.
The equations of motion are
\begin{eqnarray}
&&\nabla^{2}\psi_{1}-\frac{1}{4}\frac{\partial G(\psi_{1},\psi_{2})}{\partial\psi_{1}}F_{\mu\nu}F^{\mu\nu}-\frac{1}{2}\frac{\partial V(\psi_{1},\psi_{2})}{\partial\psi_{1}}-\frac{1}{2}\frac{\partial J_{1}(\psi_{1})}{\partial\psi_{1}}A_{\mu}A^{\mu}=0, \nonumber\\
&&\nabla^{2}\psi_{2}-\frac{1}{4}\frac{\partial G(\psi_{1},\psi_{2})}{\partial\psi_{2}}F_{\mu\nu}F^{\mu\nu}-\frac{1}{2}\frac{\partial V(\psi_{1},\psi_{2})}{\partial\psi_{2}}-\frac{1}{2}\frac{\partial J_{2}(\psi_{2})}{\partial\psi_{2}}A_{\mu}A^{\mu}=0, \nonumber\\
&&\nabla_{\mu}(G(\psi_{1},\psi_{2})F^{\mu\nu})-J_{1}(\psi_{1})A^{\nu}-J_{2}(\psi_{2})A^{\nu}=0, \nonumber \\
&&R_{\mu\nu}-\frac{1}{2}g_{\mu\nu}R=\frac{1}{2}f(\phi)(F_{\mu\alpha}F_{\nu}^{\alpha}-\frac{1}{4}g_{\mu\nu}F^{2})+\frac{1}{2}J_{1}(\psi_{1})(A_{\mu}A_{\nu}-\frac{1}{2}g_{\mu\nu}A^{2})+\frac{1}{2}J_{2}(\psi_{2})\nonumber\\
&&(A_{\mu}A_{\nu}-\frac{1}{2}g_{\mu\nu}A^{2})+\frac{1}{2}(\partial_{\mu}\psi_{1}\partial_{\nu}\psi_{1}-\frac{1}{2}g_{\mu\nu}(\partial\psi_{1})^{2})+\frac{1}{2}(\partial_{\mu}\psi_{2}\partial_{\nu}\psi_{2}-\frac{1}{2}g_{\mu\nu}(\partial\psi_{2})^{2})\nonumber\\ &&-\frac{1}{2}g_{\mu\nu}V(\psi_{1},\psi_{2}).
\end{eqnarray}
We take the fully back-reacted ansatz as
\begin{equation}
ds^{2}=-g(r)e^{-\chi(r)}dt^{2}+r^{2}(dx_{1}^{2}+dx_{2}^{2})+\frac{dr^{2}}{g(r)}, \quad \psi_{1}=\psi_{1}(r), \quad \psi_{2}=\psi_{2}(r), \quad A=\phi(r)dt.
\end{equation}

With the choice of $J_{1}=q^{2}\psi_{1}^{2}$, $J_{2}=q^{2}\psi_{2}^{2}$, and minimal coupling $\kappa_{1}=\kappa_{2}=0$, the independent equations of motion are given by~\footnote{Note that due to gauge invariance, the two scalars have the same charge.}
\begin{eqnarray}
&&\psi''_{1}+\psi'_{1}(\frac{g'}{g}-\frac{\chi'}{2}+\frac{2}{r})+\frac{q^{2}e^{\chi}\phi^{2}}{g^{2}}\psi_{1}-\frac{1}{g}(m^{2}_{1}\psi_{1}+\epsilon\psi_{2}+\eta\psi_{1}\psi^{2}_{2})=0,\nonumber\\
&&\psi''_{2}+\psi'_{2}(\frac{g'}{g}-\frac{\chi'}{2}+\frac{2}{r})+\frac{q^{2}e^{\chi}\phi^{2}}{g^{2}}\psi_{2}-\frac{1}{g}(m^{2}_{2}\psi_{2}+\epsilon\psi_{1}+\eta\psi^{2}_{1}\psi_{2})=0, \nonumber\\
&&\phi''+\phi'(\frac{\chi'}{2}+\frac{2}{r})-\frac{q^{2}(\psi^{2}_{1}+\psi^{2}_{2})}{g}\phi=0, \nonumber\\
&&\chi'+r(\psi'^{2}_{1}+\psi'^{2}_{2})+\frac{q^{2}re^{\chi}\phi^{2}}{g^{2}}(\psi^{2}_{1}+\psi^{2}_{2})=0, \nonumber\\
&&2(\psi'^{2}_{1}+\psi'^{2}_{2})+\frac{e^{\chi}\phi'^{2}}{g}+\frac{4g'}{rg}+\frac{4}{r^{2}}+\frac{-12+2m^{2}_{1}\psi^{2}_{1}+2m^{2}_{2}\psi^{2}_{2}+4\epsilon\psi_{1}\psi_{2}+2\eta\psi^{2}_{1}\psi^{2}_{2}}{g}\nonumber\\
&&+2\frac{e^{\chi}q^{2}\phi^{2}}{g^{2}}(\psi^{2}_{1}+\psi^{2}_{2})=0,\label{EOM}
\end{eqnarray}
where a prime denotes the derivative with respect to $r$, and we work in units where the AdS radius is unity.

The Hawking temperature is given by \cite{Petersen:1999zh}
\begin{eqnarray}
&&T=\frac{\sqrt{g^{rr}}}{2\pi}\frac{d}{dr}\sqrt{-g_{tt}}|_{r=r_{+}}=\frac{g'_{+}e^{-\frac{\chi_{+}}{2}}}{4\pi}\nonumber\\
&&=\frac{r_{+}}{16\pi}[(12-2m^{2}_{1}\psi^{2}_{1+}-4\epsilon\psi_{1+}\psi_{2+}-2m^{2}_{2}\psi^{2}_{2+}-2\eta\psi^{2}_{1+}\psi^{2}_{2+})e^{\frac{-\chi_{+}}{2}}-E^{2}_{+}e^{\frac{\chi_{+}}{2}}],
\end{eqnarray}
where the horizon is located at $r=r_{+}$, $E_{+}=\phi'(r_{+})$ and the subscript $+$ denotes taking the value at the horizon.

Near the boundary, the asymptotic behavior of scalar fields are in the form of
\begin{equation}
\psi_{i}=\Psi_{i}^{(1)}r^{-\Delta}+\Psi_{i}^{(2)}r^{\Delta-3}, \quad i=1,2
\end{equation}
where the renormalizable (non-renormalizable) term represents the source (expectation value) for the scalar field, and the scaling dimension of the scalar field $\Delta$ is given by
\begin{equation}
\Delta(\Delta-3)=m^{2}.
\end{equation}

In the rest of this paper, we choose $m_{1}^{2}=-2$, $m_{2}^{2}=-1$. In this choice of mass, both falloffs of scalar fields near the boundary are normalizable, and one can impose the boundary condition that either one vanishes. For simplicity, we choose the $\Psi_{i}^{(1)}$ terms to vanish, and let $\Psi_{i}^{(2)}$ be the condensates for two scalar fields~\footnote{More precisely, $\Psi_{i}^{(2)}$ corresponds to the expectation value of the scalar field operator, and the condensate is proportional to the expectation value up to some prefactor which we just neglect it.}. The condensates have the mass dimension $\lambda_{i}=3-\Delta_{i}$, where $\lambda_{1}=2$, $\lambda_{2}=\frac{3+\sqrt{5}}{2}$.

\section{Numerical study}

To solve all five independent functions ($\psi_{1}$, $\psi_{2}$, $\phi$, $g$, $\chi$), we have to impose appropriate boundary conditions at the boundary $r\rightarrow\infty$ and horizon $r=r_{h}$. At the horizon, the regularity condition is required, means $\phi(r_{h})=0$. Others can be obtained by taking Taylor expansion near the horizon and derived from the equations of motion. This leaves five undetermined parameters ($\psi_{1}(r_{h})$, $\psi_{2}(r_{h})$, $\phi'(r_{h})$, $r_{h}$, $\chi(r_{h})$). At the boundary, the five functions should behave as
\begin{eqnarray}
&&\psi_{1}=\Psi_{1}^{(1)}r^{-\Delta_{1}}+\Psi_{1}^{(2)}r^{\Delta_{1}-3}, \quad \psi_{2}=\Psi_{2}^{(1)}r^{-\Delta_{2}}+\Psi_{2}^{(2)}r^{\Delta_{2}-3}, \quad \phi=\mu-\frac{\rho}{r}, \nonumber\\
&&g=r^{2}+..., \quad \chi=0+...  \quad.
\end{eqnarray}
As discussed in the previous section, we impose the source free condition $\Psi_{i}^{(1)}=0$ since we hope the U(1) symmetry spontaneously broken. Also according to the AdS/CFT dictionary, up to a normalization, the expansion coefficients $\rho$, $\mu$, $\Psi_{i}^{(2)}\equiv\langle\mathcal{O}_i\rangle$ are interpreted as the charge density, chemical potential and condensates in the dual field theory respectively.

On the other hand, the eq.(\ref{EOM}) have the scaling symmetries
\begin{eqnarray}
&&e^{-\chi}\rightarrow\alpha^{2}e^{-\chi}, \quad \phi\rightarrow\alpha\phi, \quad t\rightarrow\alpha t,\label{scaling1}\\
&&r\rightarrow\beta r, \quad (t,x_{1},x_{2})\rightarrow\beta^{-1}(t,x_{1},x_{2}), \quad g\rightarrow \beta^{2}g, \quad \phi\rightarrow\beta\phi,
\end{eqnarray}
and one can use these two scaling symmetries to set $r_{h}=1$ and $\chi(r_{h})=0$ for performing numerics. Then we choose two of the remaining three undetermined parameters as shooting parameters to match with the source free condition $\Psi_{i}^{(1)}=0$ and solve the coupled differential equations. After solving the coupled differential equations, we need to apply the first scaling symmetry eq.(\ref{scaling1}) to set $\chi(\infty)=0$ such that the Hawking temperature can be interpreted as the temperature in the dual field theory~\cite{Hartnoll:2008kx}. Below, we fix $q=1$. We also set $\eta=0$ in our numerical calculations to focus on the effect of the Josephson coupling. We have checked that leaving the quartic scalar interaction turne on, viz. $\eta \neq 0$, we obtain similar solutions for the gauge and scalar fields, and the condensates and conductivities extracted exhibit similar behaviours, as in the $\eta = 0$ case. We leave the detailed study of the case where both the Josephson coupling and the
quartic scalar interaction are present to future work.

We emphasize here that the physical quantities of interest here are those associated with $\psi_{1,2}$. With $\eta = 0$, it is possible to go to a basis where the quadratic scalar potential becomes diagonal. However, this does not mean that the effect of the Josephson coupling is gone. The theory with respect to $\psi_{1,2}$ is not free. If one were to calculate quantities composed of $\psi_{1,2}$ (e.g. their correlation functions) in the new diagonal basis, the Josephson coupling will reappear.

\subsection{Condensates}

\begin{figure}[ht]
\centering
\subfloat[$\epsilon=0.1$]{
\label{fig:subfig:epp01}
\includegraphics[height=1.8in,width=2.5in]{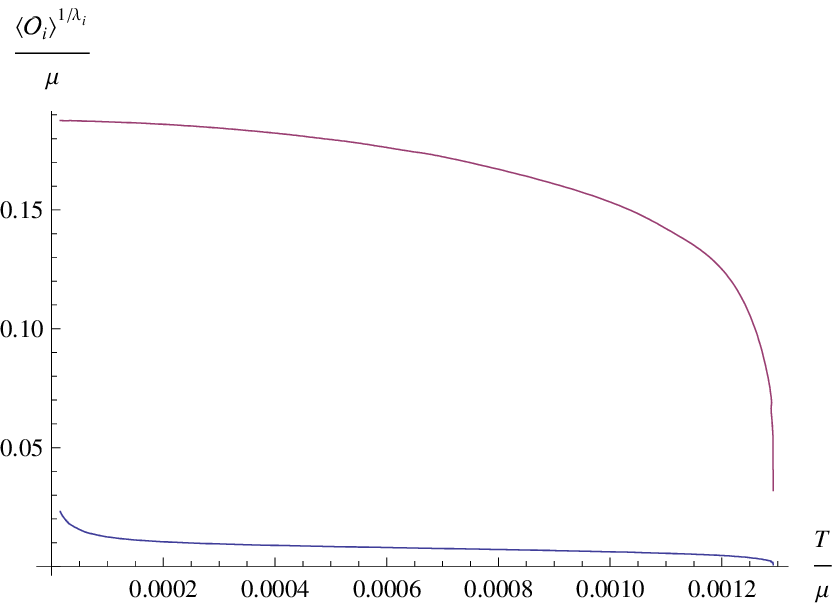}}
\hspace*{0.1in}
\subfloat[$\epsilon=-0.1$]{
\label{fig:subfig:epn01}
\includegraphics[height=1.8in,width=2.5in]{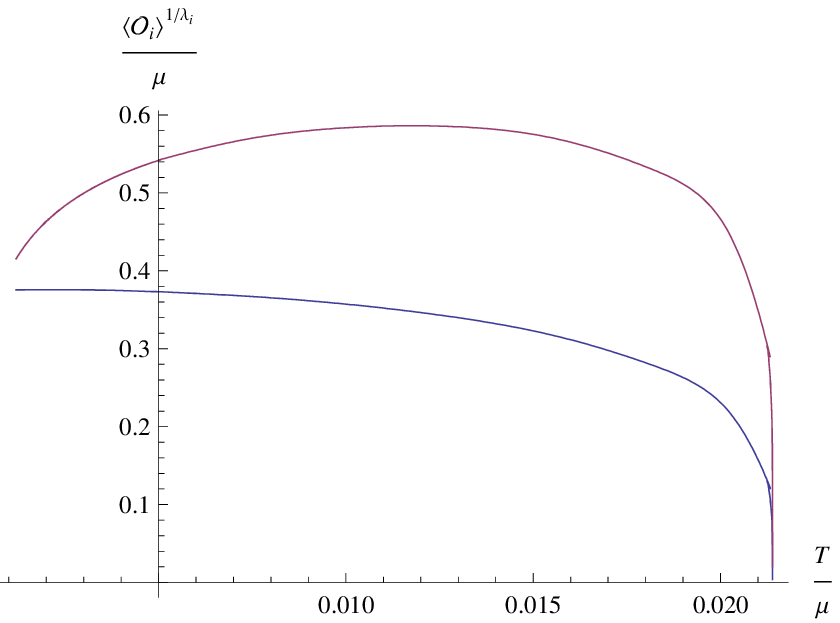}}\\
\subfloat[$\epsilon=0.5$]{
\label{fig:subfig:epp05}
\includegraphics[height=1.8in,width=2.5in]{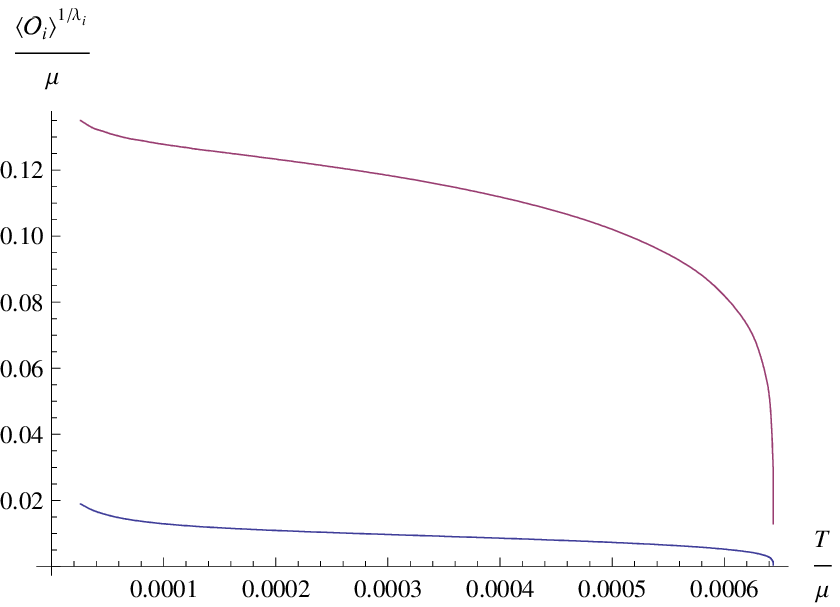}}
\hspace*{0.1in}
\subfloat[$\epsilon=-0.5$]{
\label{fig:subfig:eppn05}
\includegraphics[height=1.8in,width=2.5in]{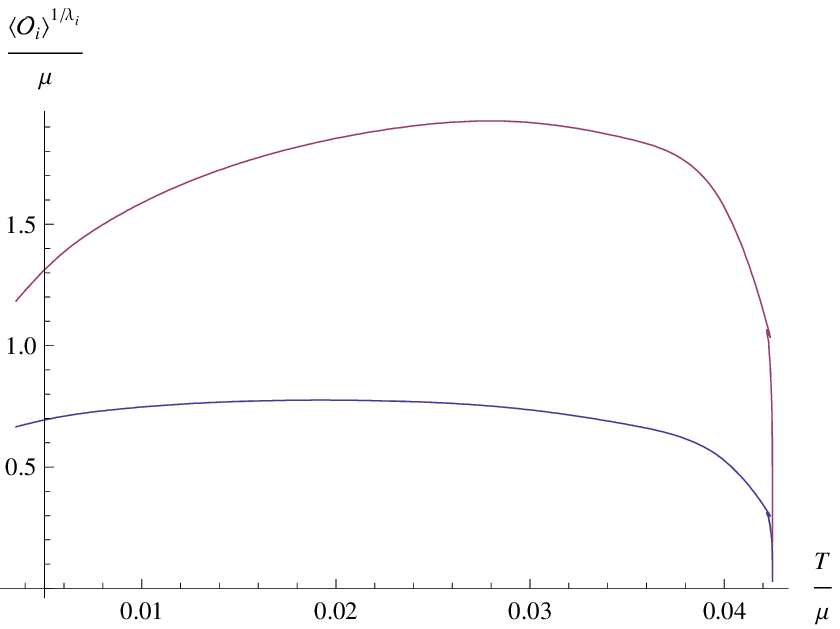}}\\
\subfloat[$\epsilon=1$]{
\label{fig:subfig:epp1}
\includegraphics[height=1.8in,width=2.5in]{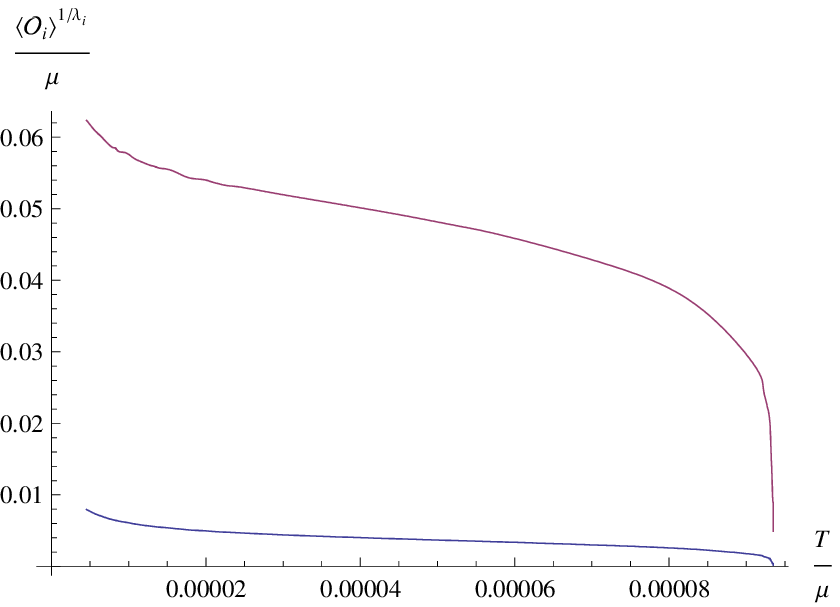}}
\hspace*{0.1in}
\subfloat[$\epsilon=-1$]{
\label{fig:subfig:epn1}
\includegraphics[height=1.8in,width=2.5in]{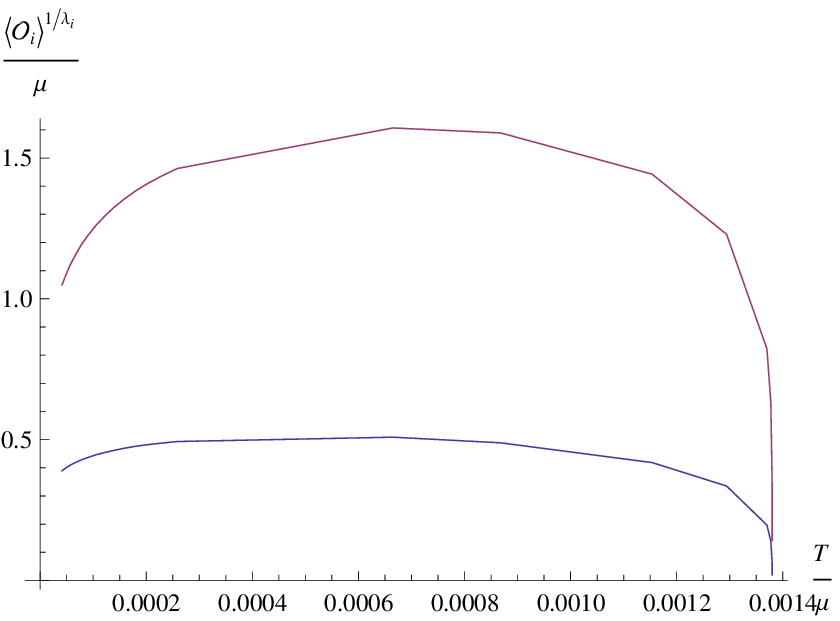}}
\caption{The two condensates, $\langle\mathcal{O}_1\rangle$ (blue) and $\langle\mathcal{O}_2\rangle$ (red), as functions of temperature for non-zero Josephson coupling, $\epsilon$. The mass dimensions of the two condensates are $\lambda_1 = 2$ and $\lambda_{2}=\frac{3+\sqrt{5}}{2}$ respectively.}
\label{fig:vevi}
\end{figure}

In Fig.~\ref{fig:vevi}, we show how the two condensates of the two charged scalar fields vary as a function of temperature. We plot the dimensionless scaling-invariant quantities $\langle\mathcal{O}_i\rangle^{1/\lambda_i}/\mu$, $i = 1,\,2$, as a function of $T/\mu$ for various values of Josephson coupling.
These scaling-invariant quantities are equivalent to those scaled to $\mu = 1$. We have checked that these hairy black hole solutions have lower free energy than the normal phase solutions without condensation, and are thus thermodynamically favored below the critical temperature.

Without the Josephson coupling, i.e. $\epsilon=0$, the critical temperature for two scalar fields are different from that found in \cite{Cai:2013wma}. When the Josephson coupling is turned on, the two scalar fields condense at the same critical temperature,
i.e. when one of the scalar condenses, it triggers the other to condense as well. This is a characteristic of two-band superconductors such as MgB$_{2}$ or Fe-based superconductors found in experiments \cite{twobandTc}. We see also that the critical temperature decreases as the strength of interband coupling increases, which is the same as the single band case.

In the weakly-coupled (BCS) theory, the value of condensate is proportional to the superconducting gap. If we n\"{a}ively extrapolate this to the strongly-coupled case here, we see interestingly from Fig.~\ref{fig:vevi} that with non-zero Josephson coupling, the ratio of two superconducting gaps in the $\epsilon>0$ case (s$_{++}$ superconductor) is higher than that in the $\epsilon<0$ case (s$_{+-}$ superconductor). If our speculation can be confirmed, this would be a novel feature predicted from our holographic model and merits further investigations, both theoretically and experimentally.

\clearpage

\subsection{Conductivity}

\subsubsection{Optical conductivity}

We are interested in the transport properties of the two-band superconductors, such as those encapsulated by the optical and thermal conductivities, which are important physical quantities measured in experiments. We compute the conductivities based on the linear response theory. Following the standard prescription in AdS/CFT correspondence, we turn on the fluctuations
$\delta A_{x}=a_{x}(r)e^{-i\omega t}$ and $\delta g_{tx}=h_{tx}(r)e^{-i\omega t}$. The fluctuation equations are given by
\begin{equation}
a''_{x}+a'_{x}(\frac{g'}{g}-\frac{\chi'}{2})+a_{x}(\frac{\omega^{2}e^{\chi}}{g^{2}}
-\frac{q^{2}(\psi^{2}_{1}+\psi^{2}_{2})}{g}) =
\frac{\phi'e^{\chi}}{g}(-h'_{tx}+\frac{2}{r}h_{tx}) \,,
\end{equation}
\begin{equation}
h'_{tx}-\frac{2}{r}h_{tx}+\phi'a_{x} = 0 \,,
\end{equation}
which can be combined into
\begin{equation}
a''_{x}+a'_{x}(\frac{g'}{g}-\frac{\chi'}{2})+[(\frac{\omega^{2}}{g^{2}}-\frac{\phi'^{2}}{g})e^{\chi}-\frac{q^{2}(\psi^{2}_{1}+\psi^{2}_{2})}{g}]a_{x}=0 \,.
\end{equation}
By solving this equation for $a_x$ with incoming wave boundary condition, the optical conductivity can be extracted from the asymptotic behavior of $a_x$ using the standard holographic prescription based on Ohm's law~\cite{Hartnoll:2009sz}:
\begin{equation}
a_x(r) = a_x^{(0)} + \frac{a_x^{(1)}}{r} + \cdots \qquad
\sigma(\omega) = \frac{J_x}{E_x} = \frac{1}{i\omega}\frac{a_x^{(1)}}{a_x^{(0)}}
\end{equation}

\begin{figure}[htbp]
\centering
\subfloat[$\mathrm{Re}\,\sigma(\omega)/\sigma_\infty$]{
\label{fig:subfig:epnReAC}
\includegraphics[width=2.95in]{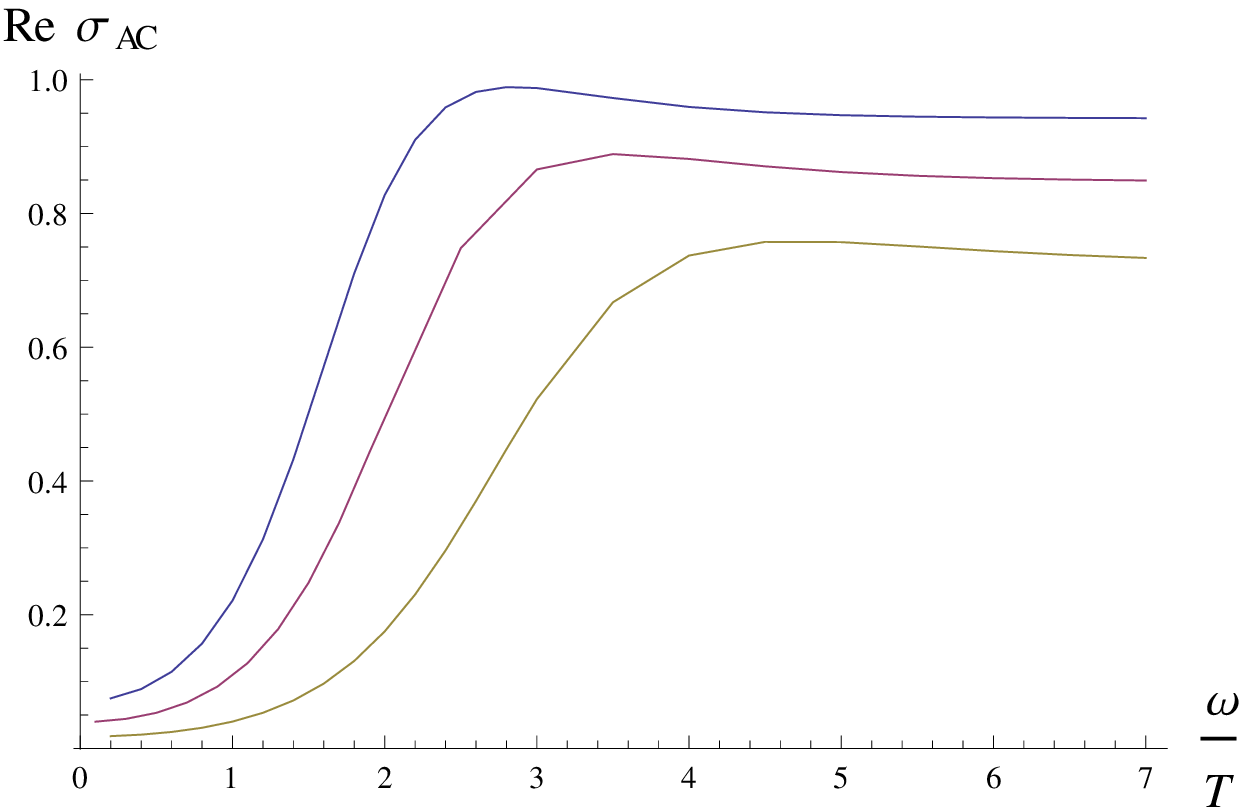}}
\hspace*{0.1in}
\subfloat[$\mathrm{Im}\,\sigma(\omega)$]{
\label{fig:subfig:epnImAC}
\includegraphics[width=3.05in]{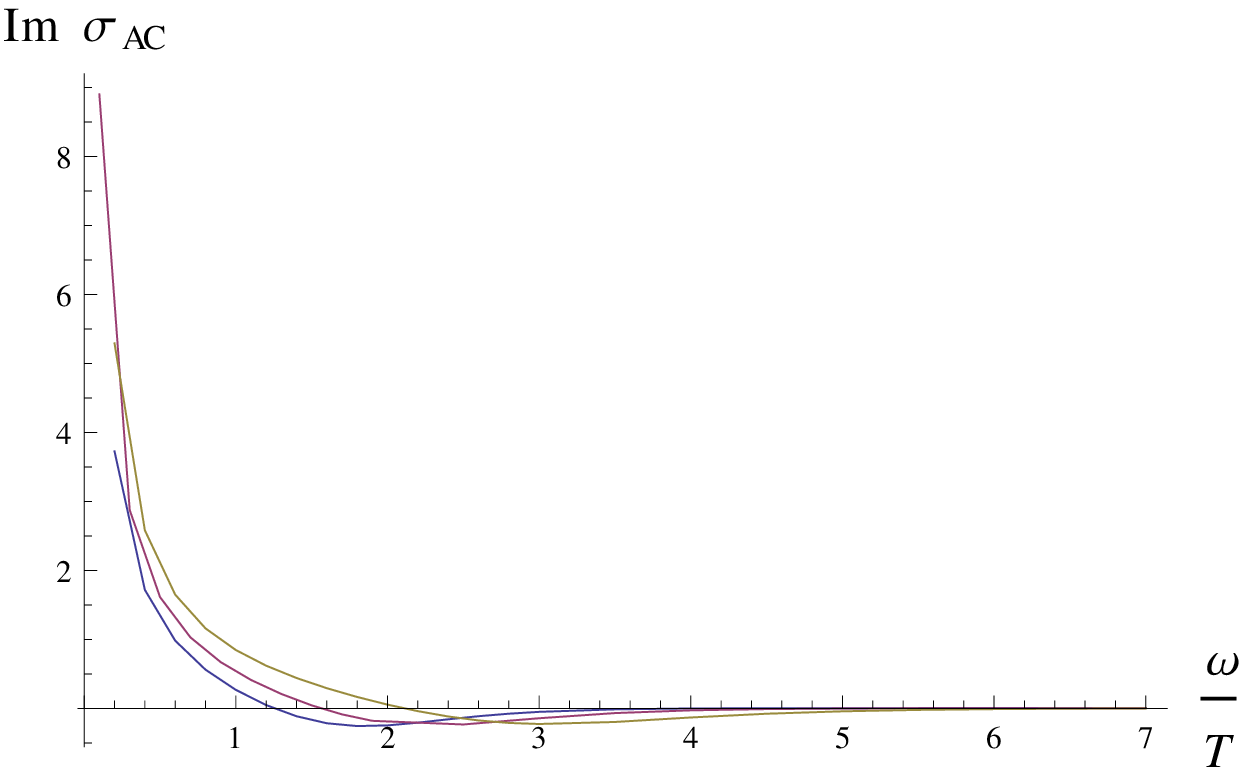}}
\caption{The AC conductivity in the case $\epsilon = -1$. The colored lines, blue, purple, brown, and green, correspond to $\mathcal{T}/\mathcal{T}_c = 0.92,\, 0.79,\,0.65,\,0.45$ respectively. In the same order,
$\sigma_\infty = 0.94,\,0.85,\,0.72,\,0.58$.}
\label{fig:sigACepn1}
\end{figure}

As an example of the typical behavior of the optical conductivity, $\sigma(\omega)$, in our model, we show in Fig.~\ref{fig:sigACepn1} the real and imaginary part of $\sigma(\omega)$ when $\epsilon = -1$ for various values of $\mathcal{T}/\mathcal{T}_c$, where we define $\mathcal{T} \equiv T/\mu$. We normalize the real part by
$\sigma_\infty\equiv\lim_{\omega\rightarrow\infty}\mathrm{Re}\,\sigma(\omega)$ to better display its features.
We see that the optical conductivity exhibits features typically seen in the one-band superconductor case. Notice the pole at $\omega = 0$ in the imaginary part of the optical conductivity. By Kramer-Kronig relation this implies a delta function in the real part of optical conductivity with the strength given by the coefficient of the pole. In our model, this coefficient does not vanish, and it approaches a constant as $T \rightarrow T_c$. This delta function at $T \geq T_c$ is due to the translational invariance, and is only visible in systems with full backreactions~\cite{Hartnoll:2008kx}. By varying the strength and sign of the interband coupling $\epsilon$, the qualitative features do not change; only $\sigma_\infty$ is changed.

\subsubsection{Thermal conductivity}

The thermal conductivity is a useful probe of nodal structure of superconductors\footnote{Other probes of the nodal structure used in experiments include specific heat, magnetic penetration length and the NMR spin lattice relaxation time}. In experiments, the low temperature behavior of the thermal conductivity is well fitted by $\kappa/T=a+bT^{\gamma-1}$, where constant part comes from the contribution of nodal excitations, while the $T^{\gamma}$ part can arise from effects that break the cooper pairs, phonons (for $\gamma=3$) or gapped excitations at low temperature.

\begin{figure}[htbp]
\centering
\subfloat[$\epsilon =$ 1 (blue), 0.5 (purple), 0.1 (brown)]{
\label{fig:subfig:eppkap}
\includegraphics[width=3in]{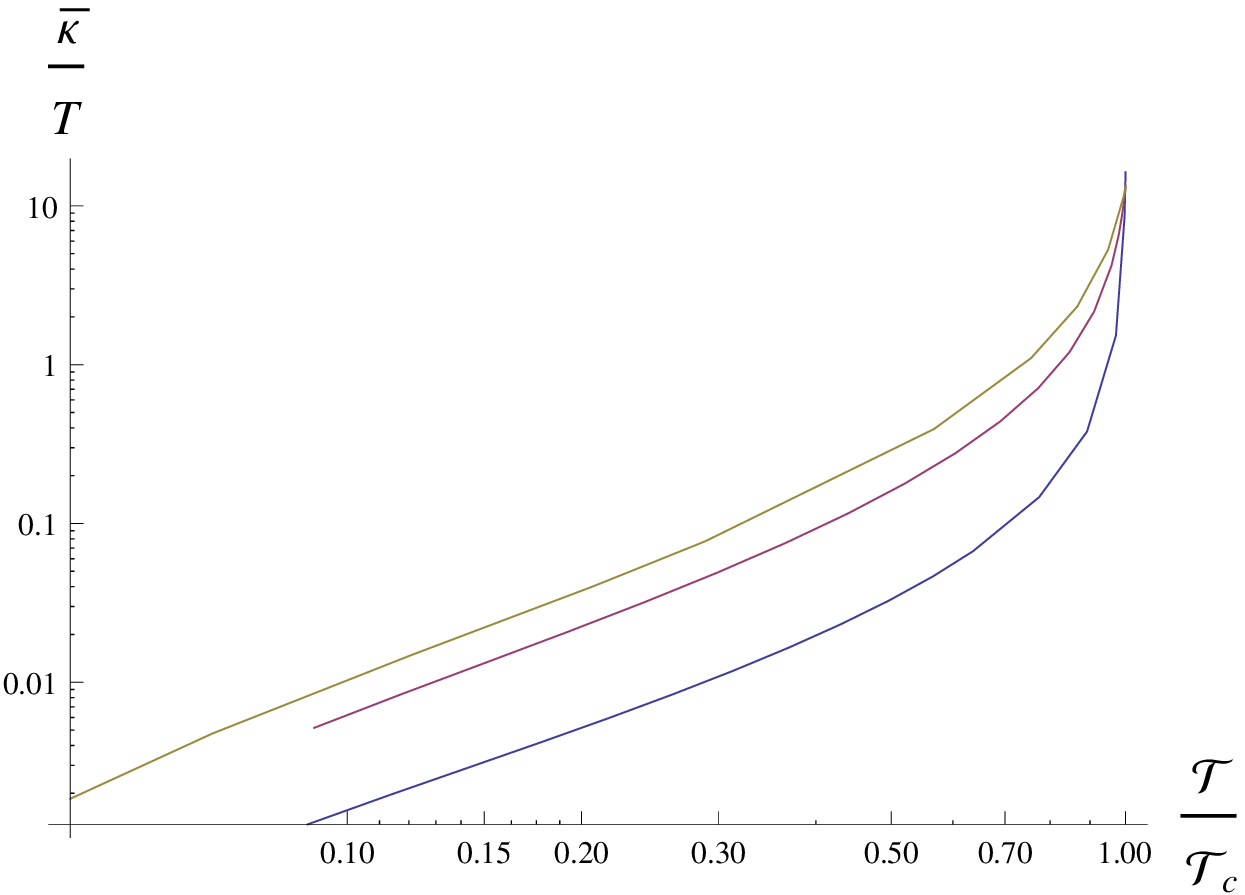}}
\hspace*{0.1in}
\subfloat[$\epsilon =$ -1 (blue), -0.5 (purple), -0.1 (brown)]{
\label{fig:subfig:epnkap}
\includegraphics[width=3in]{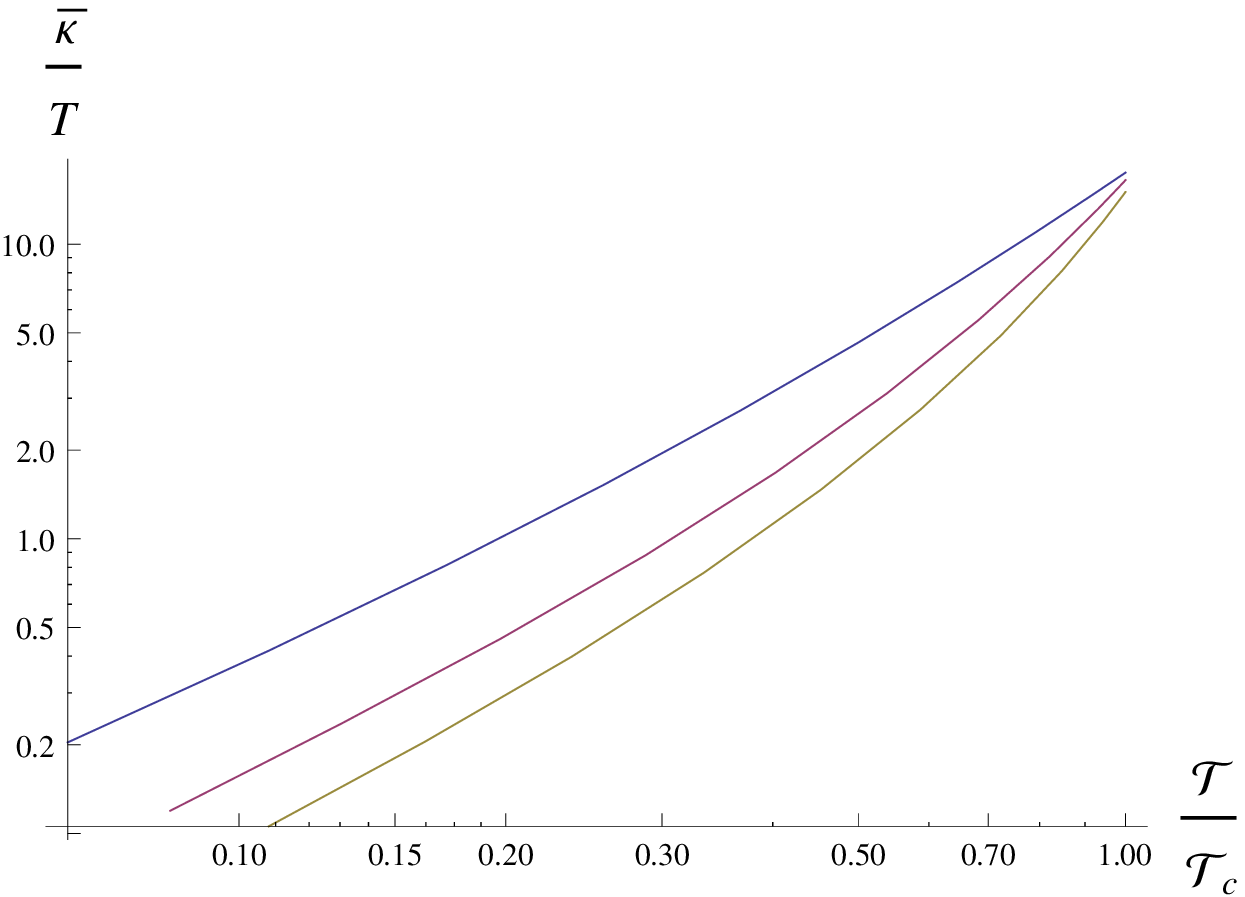}}
\caption{Thermal conductivity for various values of $\epsilon$ with both (a) positive and (b) negative signs.}
\label{fig:kap}
\end{figure}

In the holographic model, the thermal conductivity, $\bar{\kappa}(\omega)$, is given by~\cite{Hartnoll:2009sz}
\begin{equation}
T\bar{\kappa}(\omega) = \frac{i\left(\epsilon + P - 2\mu\rho\right)}{\omega} + \mu^2\sigma(\omega) \,,
\end{equation}
We see that the real part of $\bar{\kappa}(\omega)$ is determined by that of the electric conductivity alone.
In Figs.~\ref{fig:kap} we plot the behavior of
$\bar{\kappa}/T\equiv\lim_{\omega \rightarrow 0}\mathrm{Re}\,\bar{\kappa}(\omega)/T$. For convenience, we plot it as a function of
$\mathcal{T}/\mathcal{T}_c \propto T/T_c$ at various values of $\epsilon$.
At low temperature, we find for $x\equiv\mathcal{T}/\mathcal{T}_c \lesssim 0.2$, $\bar{\kappa}/T$ can be well fitted by the form
$a x^b+c x^d$, which indicates that as $T \rightarrow 0$, $\bar{\kappa}/T \rightarrow 0$ indicating the nodeless feature of our model. We list the fitted values of the parameters in Table~\ref{tab:kapfit}.
\begin{table}[h]
\caption{Values of the low temperature fit $\bar{\kappa}/T = a x^b+ c x^d$ at various $\epsilon$.}
\centering
\begin{tabular}{c|c c c c c c}
$\epsilon$ & $a$ & $b$ & $c$ & $d$ \\
\hline
1    & -0.374 & 1.238 & 0.436 & 1.274 \\
0.5  & -1.886 & 1.224 & 2.16  & 1.26 \\
0.1  & -0.139 & 1.207 & 0.758 & 1.6 \\
-0.1 & 3.562  & 1.581 & 0     & 0 \\
-0.5 & 5.36   & 1.532 & 0     & 0 \\
-1   & 3.789  & 1.152 & 11.02 & 2.011\\
\hline
\end{tabular}
\label{tab:kapfit}
\end{table}

From our fits, both the $s_{++}$ and the $s_{+-}$ states seem to be nodeless ($\bar{\kappa} \rightarrow 0$ as $T \rightarrow 0$).
For the $s_{++}$ state, this is to be expected due to the existence of the superconducting gap, as is confirmed by experiments \cite{nodeless}. The situation for the $s_{+-}$ state is less clear experimentally. The $s_{+-}$ state is widely believed to appear in iron-based superconductors. While most families of iron-based $s_{+-}$ superconductors are found to be nodeless, not all of them are. Some families such as LaFePO are found to have a residual linear temperature dependence in $\kappa$ at low temperature, and there is nodal excitations in at least one of its bands \cite{iron-based3}.

Experimentally, the thermal conductivity of a fully gapped (and thus nodeless) superconductor is seen to have at least a power-law temperature dependence at low temperature with an exponent larger than three. In the case with nodes however, the power-law exponent can be arbitrary depending on how the cooper pairs are broken. In our holographic model, we found the power-law exponent to be less than three for all the values of Josephson coupling we looked at. This maybe a feature of our holographic model, which requires further investigation beyond the scope of the present work. But given that confusions remain under what circumstances $s_{+-}$ superconductors are nodeless experimentally, we caution against too literal a comparison with current experiments.

From Fig.~\ref{fig:kap}, we find the temperature dependence of $\bar{\kappa}/T$ in $s_{+-}$ and $s_{++}$ states are quite different near the critical temperature. We see that the thermal conductivity increases faster for a holographic $s_{++}$ superconductor than an $s_{+-}$ one
as the temperature increases. This might explain the result that the critical temperature of the $s_{+-}$ state is generically higher than the $s_{++}$ state (for the normalized critical temperature $T^{nor}\equiv T_{c}/\mu_{c}$ at various $\epsilon$, see Table \ref{tab:Tc}): if the $s_{++}$ state is more susceptible to thermal excitations than the $s_{+-}$ state, the cooper pairs in the $s_{++}$ state would be easier to break than in the $s_{+-}$ state as the temperature increases, resulting in an exit from superconductivity at a lower temperature.

\begin{table}[h]
\caption{The normalized critical temperature $T^{nor}_{c}\equiv T_{c}/\mu_{c}$ at various $\epsilon$.}
\centering
\begin{tabular}{c|c c|c c|c c}
$\epsilon$ & 1 & -1 & 0.5 & -0.5 & 0.1 & -0.1 \\
\hline
$T^{nor}_{c}$ & $9.35\times 10^{-5}$ & $1.38\times 10^{-3}$ & $6.44\times 10^{-4}$ & $4.25\times 10^{-2}$ & $1.29\times 10^{-3}$ & $2.14\times 10^{-2}$\\
\hline
\end{tabular}
\label{tab:Tc}
\end{table}

\section{Analytic study}

\begin{figure}
\center{
\includegraphics[width=8cm]{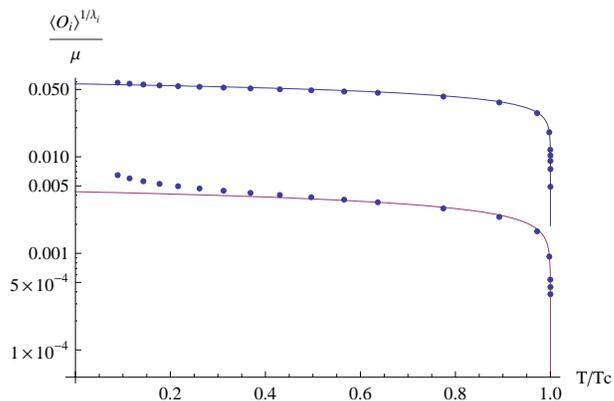}\hspace{0.2cm}
\caption{\label{fig:condensate_analytic} Analtyic fit (curves) of two condnsates near and below the critical temperature agrees well with numerical results (dots).  We remark that to obtain the analytic result, we have adopted oupling $\epsilon=1$ and ${\cal O}_{12}=36$.}
}
\end{figure}

In previous sections, we have investigated the two-band model numerically.  It would be also insightful to study the connection between condensates and other variables in the model via some analytic method.  Many analytic approaches have been proposed to address the universal properties of second order phase transitions in holographic superconductors\cite{Ge:2010aa, Siopsis:2010uq, Zeng:2010zn,Li:2011xja,Cai:2011ky,Chen:2011en,Ge:2011cw,Momeni:2011iw}. In particular, it would be interesting to apply the variational method for the Sturm-Liouville eigenvalue problem in \cite{Siopsis:2010uq}, to our two-band model.  First we notice that near the critical temperature, where one can neglect the backreaction, the $\Phi$ and $\Psi_i$ takes the following forms:
\begin{eqnarray}
&&\Phi = \lambda r_+ (1-z),\nonumber\\
&&\Psi_i = \frac{<{\cal O}_i>}{\sqrt{2}r_+^{\Delta_i}}z^{\Delta_i}F^i(z),
\end{eqnarray}
where $\lambda = \frac{\rho}{r_{+c}^2}$ and $\Delta_i^{\pm}=\frac{3}{2}\pm \sqrt{\frac{9}{4}+m_i^2}$.
Applying Sturm-Liouville theorem to the equations of condensate fields, one can minimize the eigenvalue $\lambda^2$ providing the coupling $\epsilon$ and condensate ratio ${\cal O}_{12}\equiv \frac{<{\cal O}_1>}{<{\cal O}_{2}>}$ at the critical temperature.  To be specific, we have to minimize
\begin{align}\label{analytic-eqn}
\lambda^2 &= \frac{1}{\int\!dz\left[W_1(z)F_1(z)^2+W_2(z)F_2(z)^2\right]}
\bigg\{\int\!dz\left[P_1(z)F'_1(z){}^2 + P_2(z)F'_2(z){}^2\right] \notag \\
&\quad +\int\!dz\left\{\left[Q_1(z)+R_1(z)\right]F_1(z)^2 + \left[Q_2(z)+R_2(z)\right]F_2(z)^2\right\}
\bigg\}
\end{align}
for trial functions $F_i=1-\alpha_i z^2$.  The functions $P_i(z),Q_i(z),R_i(z)$ are derived in the Appendix.  Then one can read the critical temperature as a function of $\rho$:
\begin{equation}
T_c = \frac{3}{4\pi}\sqrt{\frac{\rho}{\lambda_m}}
\end{equation}

To compare with our numerical results, here we focus on the same choice for the conformal dimensions of condensates.  Following similar derivation in \cite{Siopsis:2010uq}, one can express the condensates near and below the critical temperature as follows:
\begin{eqnarray}\label{eq:condensate_fit}
&&<{\cal O}_1> \simeq (1-\frac{T}{T_c})^{1/2}\gamma_1 T_c^{\Delta_1} (1+(\frac{\gamma_1}{\gamma_2})^2 T_c^{2(\Delta_1-\Delta_2)}{\cal O}^{-1}_{12})^{-1/2},\nonumber\\
&&<{\cal O}_2>\simeq (1-\frac{T}{T_c})^{1/2}\gamma_2 T_c^{\Delta_2} (1+(\frac{\gamma_2}{\gamma_1})^2 T_c^{2(\Delta_2-\Delta_1)}{\cal O}_{12})^{-1/2},
\end{eqnarray}
where $\gamma_i \equiv \frac{2}{\sqrt{C_i}}(\frac{4\pi}{3})^{\Delta_i}$ and $C_i=\int{\lambda m_i^2 \frac{z^{2(\Delta_i-1)}(1-z)}{1-z^3}}F_i^2 dz$.  In the figure (\ref{fig:condensate_analytic}), we showed that the analytical approximation (\ref{eq:condensate_fit}) agrees very well with the numerical results near the critical point.

\begin{figure}
\center{
\includegraphics[width=8 cm]{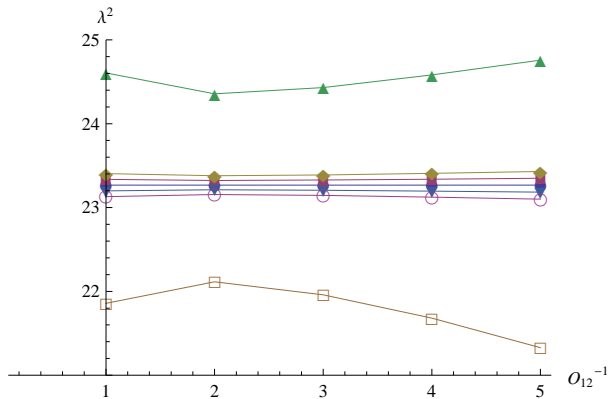}\hspace{0.2cm}
\caption{\label{fig:parameter_Tc}  The eigenvalues $\lambda^2$ (vertical axis) is plotted against the condensate ratio (horizontal axis) at the $T_c$.  Different curves, from top to down, correspond to $\epsilon = 1,0.1, 0.05, 0, -0.05, -0.1, -1$.}
}
\end{figure}

The analytic method also has the advantages to easily reveal the connection between model parameters.  In the figure (\ref{fig:parameter_Tc}), we plot the fitting curve of eigenvalues $\lambda^2$ as a function of $\epsilon$ and condensate ratio ${\cal O}_{12}$.  We conclude that the $T_c$ slightly decreases (increases) for positive (negative) coupling and remains nearly same for different condensate ratios at small coupling.

\section{Summary and Outlooks}

We have constructed a fully back-reacted holographic model of two-band superconductor with an explicit interband coupling between the two charged scalars. The sign of the interband coupling indicates whether the pairings of two bands is in phase or out of phase. We have studied its effects on the two condensates and the critical temperature. We have shown that in the presence of the interband coupling, when one scalar field condenses, it will induce the other scalar to condense at the same critical temperature, and the critical temperature decreases as the strength of the interband coupling increases. The ratio of the two gaps in the $s_{++}$ state is larger than in the $s_{+-}$ state, but the critical temperature of $s_{+-}$ is generically higher~\footnote{The critical temperature of the $s_{+-}$ state larger than that of the $s_{++}$ state is consistent with earlier studies~\cite{Ummarino:2009,Krikun:2012yj}}.

We have also studied the transport properties of the holographic two-band superconductor, and we calculated its optical and thermal conductivities. The optical conductivity is qualitative similar to that of the single band superconductor, while the thermal conductivity seems to indicate that our model has no nodal excitations. Our study is primarily a numerical one. But in regimes where the Sturm-Liouville method is applicable, analytically results can be obtained, and is fully consistent with our numerical results.

There are many directions for future works. One is to see in the higher frequency region of the optical conductivity if there exists a mid-infrared peak when the interband coupling is large. In this work, we worked with a translational invariant system with no impurities. It would be interesting to introduce impurities in our model, as the mid-infrared peak in the optical conductivity is expected when the scattering between the impurities and charge carriers is large. Furthermore, it would be interesting to study the impurity induced $s_{+-} \rightarrow s_{++}$ transition as discussed in Ref.~\cite{Efremov:2012} in our model.

To be completely sure of the nodal structure, the strict zero temperature limit should be taken in our model. As was shown in the
single band case~\cite{Horowitz:2009ij}, the bulk geometry could be quite different when $T$ is strictly zero from when $T$ is small but nevertheless finite. It is reasonable to expect this applies to the two-band case as well, and new solutions for the strict $T = 0$ case have to be found. Another way to probe the gap structure of the superconductor complementary to the thermal conductivity is to study the specific heat. A generalization of the discussions in Ref.~\cite{Hartnoll:2012pp} to the two-band case would be an immediate next step.

The strongest indication for an $s_{+-}$ superconductor is in the neutron spin measurement, where there is a resonance peak in the dynamical spin susceptibility at $\omega \sim 2\Delta$~\cite{spinresonance1,spinresonance2}~\footnote{Note such spin resonances are only found in unconventional superconductors such as the $s_{+-}$ or the d-wave superconductors.}. If we can see this feature in our model, we can be sure that at negative Josephson coupling we are indeed modeling the $s_{+-}$ superconductor.

The response of the two-band superconductor to an external magnetic field presents many very interesting questions. For one, magnetic field can significantly change the temperature dependence of the thermal conductivity, and it would be very interesting to see what it would be in our model.

It has been argued that the $s_{++}$ superconductor could be the so-called ``type-1.5'' superconductor~\cite{phase-locking,type 1.5}, which has the unusual properties that the intervortex interaction is attractive at long range and repulsive at short range \cite{Carlstrom:2010,semi-Meissner1,Geurts:2010,Chave:2011,Silaev:2011}, and vortex clusters coexisting with Meissner domain at intermediate field strength forming the so-called "semi-Meissner" state~\cite{semi-Meissner1,semi-Meissner2}. More technically, for a two-band type-1.5 superconductor, the coherence lengths for the two bands, $\xi_{1}$ and $\xi_{2}$, and the magnetic penetration length, $\lambda$, satisfy the relation $\xi_{1}<\sqrt{2}\lambda<\xi_{2}$ \cite{Carlstrom:2010,Silaev:2011,semi-Meissner1,Komendova:2011,Komendova:2012}.
As steps to confirm whether the type-1.5 state truly exists, it would be very interesting to verify this relation, and to look for a first order phase transition between the Meissner and the semi-Meissner state.

\appendix

\section{Derivation of equation (\ref{analytic-eqn})}

In this appendix, we give a derivation of Eq.~(\ref{analytic-eqn}). Near the critical temperature, the equations of motion for scalar fields are simplified as those in the probed limit:
\begin{eqnarray}\label{probe-eqn}
&&\Psi_1^{''}+\frac{f^\prime}{f}\Psi_1^{\prime}+\frac{r_+^2}{z^4}(\frac{\Phi^2}{f^2})\Psi_1-\frac{\epsilon r_+^2}{z^4 f}\Psi_2 = 0,\nonumber\\
&&\Psi_2^{''}+\frac{f^\prime}{f}{\Psi_2}^{\prime}+\frac{r_+^2}{z^4}(\frac{\Phi^2}{f^2})\Psi_2-\frac{\epsilon r_+^2}{z^4 f}\Psi_1 = 0,
\end{eqnarray}
where we have defined $z=r_+/r$ and derivative $\prime$ respects to $z$.  The leading order terms in the fields $\Phi(z)$ and $\Psi_i(z)$ take following forms:
\begin{eqnarray}
&&\Phi(z) = \lambda r_+ (1-z),\nonumber\\
&&\Psi_i(z) = \frac{<{\cal O}_i>}{\sqrt{2}r_+^{\Delta_i}}F_i(z),
\end{eqnarray}
for some trial functions $F_i(z)$.  Substitute them into (\ref{probe-eqn}) and one obtains
\begin{eqnarray}
&&F_1^{''}+(\frac{f^{\prime}}{f}+\frac{2\Delta_1}{z})F_1^{\prime}+(\frac{\Delta_1(\Delta_1-1)}{z^2}+\frac{f^{\prime}}{f}\frac{\Delta_1}{z}-\frac{r_+^2}{z^4}\frac{m_1^2}{f}+\lambda^2\frac{r_+^4(1-z)^2}{z^4f^2})F_1(z)\nonumber\\
&&-\epsilon\frac{r_+^{\Delta_1-\Delta_2+2}}{z^{\Delta_1-\Delta_2+4}f}{\cal O}_{12}F_2=0,\nonumber\\
&&F_2^{''}+(\frac{f^{\prime}}{f}+\frac{2\Delta_2}{z})F_2^{\prime}+(\frac{\Delta_2(\Delta_2-1)}{z^2}+\frac{f^{\prime}}{f}\frac{\Delta_2}{z}-\frac{r_+^2}{z^4}\frac{m_2^2}{f}+\lambda^2\frac{r_+^4(1-z)^2}{z^4f^2})F_2(z)\nonumber\\
&&-\epsilon\frac{r_+^{\Delta_2-\Delta_1+2}}{z^{\Delta_2-\Delta_1+4}f}{\cal O}_{12}^{-1}F_1=0.
\end{eqnarray}
Multiply each equation with $z^{2\Delta_i}f$ respectively, one can further put them in the following form:
\begin{eqnarray}
&&-[P_1F_1^{\prime}]^{\prime}+Q_1F_1+R_2F_2=\lambda^2 W_1F_1,\nonumber\\
&&-[P_2F_2^{\prime}]^{\prime}+Q_2F_2+R_1F_1=\lambda^2 W_2F_2,
\end{eqnarray}
with
\begin{eqnarray}
&&P_i=z^{2\Delta_i}f, \quad Q_i = -\Delta_i (\Delta_i-1) z^{2\Delta_i-2}f-\Delta_i z^{2\Delta_i-1}f' + m_i^2 r_+^2 z^{2\Delta_i-4},\quad W_i = r_+^2 z^{2\Delta_i-4}(1-z)^2f^{-1},\nonumber\\
&&R_1 = \epsilon r_+^{\Delta_2-\Delta_1+2}{\cal O}_{12}z^{\Delta_1+\Delta_2-4}, \quad R_2 = \epsilon r_+^{\Delta_1-\Delta_2+2}{\cal O}_{12}^{-1}z^{\Delta_1+\Delta_2-4}.
\end{eqnarray}
If one is able to find a common eigenvalue $\lambda$ to minimize both equations as follows:
\begin{eqnarray}
&&\int{P_1F^{\prime}_1{}^2}dz+\int{(Q_1F_1^2+R_2F_1F_2)}dz = \lambda^2 \int{W_1F_1^2}dz,\nonumber\\
&&\int{P_2F^{\prime}_2{}^2}dz+\int{(Q_2F_2^2+R_1F_1F_2)}dz = \lambda^2 \int{W_2F_2^2}dz,
\end{eqnarray}
where each equation is minimized according to the variation method of Sturm-Lioville theorem and integration is from $z=0$ to $1$.  Then the same eigenvalue can surely minimize the sum of them, say
\begin{equation}
\int{(P_1F^{\prime}_1{}^2+P_2F^{\prime}_2{}^2)}dz+\int{[(Q_1+R_1)F_1^2+(Q_2+R_2)F_2^2]}dz=\lambda^2\int{(W_1F_1^2+W_2F_2^2})dz.
\end{equation}
In other words, we can express $\lambda^2$ as in the equation (\ref{analytic-eqn}).  We remark this expression reduces to the variation of single field as in \cite{Siopsis:2010uq} if coupling $\epsilon$ and  one of the field $\Psi_i$ are swtiched off.
\begin{acknowledgments}
The authors would like to thank Jiunn-Yuan Lin and Wei-Feng Tsai for many valuable discussions, comments and useful references. The authors also thank to Egor Babaev and Miload Milo\v{s}evi\'{c} for bringing us many useful references and correct the citations.  WYW was grateful to the hospitality of YITP while preparing the draft. This work was supported in parts by the National Science
Council under grants NSC 101-2811-M-009-015, NSC 102-2811-M-009-057 (SYW), NSC 100-2112-M-033-009-MY2, NSC 102-2112-M-033-003-MY4 (WYW), and the National Center for Theoretical Science, Taiwan.
\end{acknowledgments}

\bibliography{apssamp}

\end{document}